\documentclass[seceq,preprint]{ptptex}

\usepackage{graphicx}
\def\fsl#1{\setbox0=\hbox{$#1$}                 
   \dimen0=\wd0                                 
   \setbox1=\hbox{/} \dimen1=\wd1               
   \ifdim\dimen0>\dimen1                        
      \rlap{\hbox to \dimen0{\hfil/\hfil}}      
      #1                                        
   \else                                        
      \rlap{\hbox to \dimen1{\hfil$#1$\hfil}}   
      /                                         
   \fi}                                         %

\notypesetlogo                       
\preprintnumber[3cm]{
UWO-TH-07/06}{
\markboth{
V. A. Miransky%
}{
Dynamics with Vector Condensates%
}

\title{
Dynamics with Vector Condensates at Finite Density in QCD and
Beyond%
}

\author{
V. A. \textsc{Miransky}\footnote{e-mail address:
vmiransk@uwo.ca}} 

\inst{
Department of Applied Mathematics, University of Western 
Ontario, London, Ontario N6A 5B7, Canada\\
and\\ 
Yukawa Institute for Theoretical Physics,
Kyoto University, Kyoto 606--8502, Japan\\}

\abst{
I describe the dynamics in recently revealed phases with vector
condensates of gauge fields in dense QCD (gluonic phase)
and other gauge
models. In this case, the Higgs mechanism is
provided by condensates of gauge (or gauge plus scalar) fields.
Because most of the initial symmetries
in such systems are spontaneously broken, their dynamics is very rich.
In particular, by using the Ginzburg-Landau approach, 
the existence of a gluonic phase with
both the rotational symmetry and the electromagnetic
$U(1)$ being spontaneously broken was established. In other words, this phase 
describes an anisotropic superconductor. 
In the dual (confinement) description of this dynamics
in dense two-flavor QCD, there are light exotic vector hadrons in the 
spectrum, some of which condense. There are also vortex-like and 
roton-like excitations in these phases.}

\begin{document}

\maketitle


\section{Introduction}
\label{sec1}

In this review, I summarize results obtained by
our group in the description of the
dynamics in recently revealed
phases with vector condensates of gauge fields in dense 
two-flavor QCD
(gluonic phase) 
\cite{Gorbar:2005rx,Gorbar:2005tx,Gorbar:2006up,Hashimoto:2006mn,Gorbar:2007vx}
and in other gauge models.\cite{sigmamodel,Gorbar:2005pi,Buchel:2006aa,Buchel:2007rx}
Let me at once emphasize that the vector condensates of gluon fields
in dense QCD are very different from the conventional gluon
condensate in vacuum QCD. \cite{SVZ} While the latter is
Lorentz invariant, the former are not. In fact,
since vacuum expectation values of spatial components of vector
fields break
the rotational symmetry, it is natural to have a spontaneous
breakdown both of external and internal symmetries in this
case. Dynamics in such systems are very rich.

It is expected that at sufficiently high baryon density, cold quark
matter should be in a color superconducting state (for reviews, 
see Ref. \citen{review}). 
On the other hand, it was suggested long ago that quark matter might
exist inside the central region of compact stars.\cite{quark_star}
This is one of the main reasons why the dynamics of
the color superconductivity 
has been intensively studied.  

Bulk matter in compact stars should be in $\beta$-equilibrium,
providing by weak interactions, 
and be electrically and color neutral.
The electric and color neutrality conditions play a crucial role
in the dynamics of quark pairing.
\cite{Iida:2000ha,Alford:2002kj,
Huang:2003xd,Abuki:2004zk,Ruster:2005jc}. 
Also, in the dense quark matter,
the strange quark mass cannot be neglected.
These factors lead
to a mismatch $\delta\mu$ between the Fermi momenta of
the pairing quarks.

As was revealed in Refs. \citen{Huang:2004bg,Huang:2004am},
the 
gapped (2SC) and gapless (g2SC)
two-flavor color superconducting phases \cite{Huang:2003xd}
suffer from a chromomagnetic
instability connected with the presence of imaginary 
Meissner masses of gluons.
While the 8th gluon has an imaginary Meissner mass only
in the g2SC phase, with the diquark gap
$\bar{\Delta} < \delta\mu$ (an intermediate coupling regime),
the chromomagnetic instability for the 4-7th gluons appears also in
a strong coupling regime, with
$\delta\mu < \bar{\Delta} < \sqrt{2}\delta\mu$.
Later a chromomagnetic instability was also found in
the three-flavor gapless color-flavor locked (gCFL)
phase.\cite{Casalbuoni:2004tb}.

Meissner and Debye masses are screening (and not pole) ones.
It has been recently revealed in Ref. \citen{Gorbar:2006up}
that the chromomagnetic instabilities 
in the 4-7th and 8th gluonic channels correspond to two very
different tachyonic spectra of plasmons.
It is noticeable that while
(unlike the Meissner mass)
the (screening) Debye mass for an electric mode remains real for
all values of $\delta\mu$ both in the 2SC and g2SC phases 
\cite{Huang:2004bg,Huang:2004am}, 
the tachyonic plasmons occur both for the magnetic and electric modes
\cite{Gorbar:2006up}. The
latter is important since it clearly shows that this instability is connected
with vectorlike excitations: Recall that
two magnetic modes correspond to two 
transverse components of a plasmon, and one electric mode
corresponds to its longitudinal component. 
This form of the plasmon spectrum leads to the unequivocal conclusion
about the existence of vector condensates of gluons
in the ground state 
of two flavor quark matter with 
$\bar{\Delta} < \sqrt{2}\delta\mu$, thus
supporting the scenario with gluon condensates (gluonic phase)
proposed in Ref. \citen{Gorbar:2005rx}. 
While the analysis in paper \citen{Gorbar:2005rx} was done only in
the vicinity of the critical point 
$\delta\mu \simeq \bar{\Delta}/\sqrt{2}$,
a numerical analysis of the gluonic phase far away of the scaling
region ~was considered in Refs.
\citen{Fukushima:2006su,Kiriyama:2006ui}. It confirms
the general picture suggested in  Ref.~\citen{Gorbar:2005rx}.

\section{Renormalizable model for dynamics with vector condensates}
\label{sec2}

Since a dynamics with vector condensates is a rather new ``territory'',
it would be important to have an essentially soluble model which
would play the same role for such a dynamics
as the linear $\sigma$
models play for the conventional dynamics with spontaneous symmetry
breaking with condensates of scalar fields. Fortunately, such a model
exists: it is the gauged linear $SU(2)_L\times U(1)_Y$
$\sigma$-model (without fermions) with a chemical potential for
hypercharge $Y$ considered 
in Ref. \citen{sigmamodel}. Let me describe this
model. It will be very useful for better understanding the dynamics in 
the gluonic phase.

\subsection{Vacuum solution as anisotropic superconducting medium}
\label{sec3}
 
The Lagrangian density of this model reads (the metric 
$g^{\mu\nu} = \mbox{diag}(1, -1, -1, -1)$ is used):
\begin{equation} 
\begin{split}
{\cal L}=-\frac{1}{4}F^{(a)}_{\mu\nu}F^{\mu\nu(a)}-
\frac{1}{4}F^{(Y)}_{\mu\nu}F^{\mu\nu(Y)} +
[(D_{\nu}-i\mu_Y\delta_{\nu0})\Phi]^{\dag}
(D^{\nu}-i\mu_Y\delta^{\nu0})\Phi\\
-m^2\Phi^{\dag}
\Phi-\lambda(\Phi^{\dag}\Phi)^2,
\end{split}
\label{Lagrangian}
\end{equation}
where the covariant derivative $D_{\mu}=\partial_{\mu}-igA_{\mu}-
(ig^{\prime}/2)B_{\mu}$, $\Phi$ is a complex doublet field
$\Phi^T=(\varphi^+,\varphi_0)$, and the chemical
potential $\mu_Y$ is provided by external conditions (to be specific, we take
$\mu_Y >0$). Here
$A_{\mu}=A_{\mu}^{(a)}\tau^a/2$ are $SU(2)_L$ gauge fields ($\tau^a$ are 
three Pauli matrices) and the field strength
$F^{(a)}_{\mu\nu}=\partial_{\mu}A_{\nu}^{(a)}-
\partial_{\nu}A^{(a)}_{\mu} + g\epsilon^{abc}A^{(b)}_{\mu}A^{(c)}_{\nu}$.
$B_{\mu}$ is a $U_{Y}(1)$ 
gauge field with the field strength $F^{(Y)}_{\mu\nu}  =
\partial_{\mu}B_{\nu}-\partial_{\nu}B_{\mu}$. The hypercharge of the doublet
$\Phi$ equals +1. This model has the same
structure as the electroweak theory without fermions and with the chemical
potential for hypercharge $Y$. 
Henceforth
we will omit the subscript $L$, allowing various
interpretations of the $SU(2)$.
\footnote{Note that because the $U(1)_Y$ symmetry
is local, for a nonzero chemical potential $\mu_Y$
one should introduce a source term $B_0J_0$ in Lagrangian density
(\ref{Lagrangian}) in order to make the system neutral with respect to
hypercharge $Y$ (the Gauss law). 
The value of the background hypercharge density $J_0$
(representing heavy particles) is determined from the requirement that
$B_0=0$ is a solution of the equation of motion  
for $B_0$.\cite{sigmamodel} There exists an alternative description of
this dynamics in which a background hypercharge density $J_0$ is
considered as a free parameter and $\mu_Y$ is taken to be zero. 
Then the Gauss law will define the vacuum expectation value 
$\langle{B_0}\rangle$. It is not difficult to check that these two approaches
are equivalent if the chemical potential $\mu_Y$ in the first 
approach is taken to be equal to the value
$\frac{g'}{2}\langle{B_0}\rangle$ from
the second one.}   

The model is renormalizable and for small coupling 
constants $g$, $g^{\prime}$ and $\lambda$, the tree approximation 
is reliable there.
Because the chemical potential explicitly breaks
the Lorentz symmetry, the symmetry of the model is 
$SU(2) \times U(1)_Y \times SO(3)_{\rm rot}$.
As was shown in Ref. \citen{sigmamodel}, for sufficiently large
values of the chemical potential $\mu_Y$, the condensates of
both the scalar doublet $\Phi$ {\it and} the gauge field $A_\mu$ occur. 
The ground state solution is given by
\begin{equation}
\langle{\Phi^T}\rangle =(0,v_0), \quad
|\langle{W^{(-)}_z}\rangle|^2 \equiv |C|^2
= \frac{\mu_Y v_0}{\sqrt{2}g}-\frac{v_0^2}{4},
\quad \langle{A^{(3)}_0}\rangle \equiv D
= \frac{v_0}{\sqrt{2}}, 
\label{vacuum}
\end{equation}
where
\begin{equation}
v_0=\frac{\sqrt{(g^2+64\lambda)\mu_{Y}^2-
8(8\lambda-g^2)m^2}-3g\mu_Y}{\sqrt{2}
(8\lambda-g^2)} \,,
\label{v0}
\end{equation}
$W^{(\mp)}_{\mu}=\frac{1}{\sqrt{2}}(A_{\mu}^{(1)} \pm iA_{\mu}^{(2)})$,
$\Phi^T=(\varphi^+,\varphi_0)$,
and the vacuum expectation values of 
all other fields are equal to zero.
\footnote{Here ``sufficiently large values of
$\mu_Y$'' means the following: When $m^2 \geq 0$, $\mu_Y$ should be larger
than the critical value $\mu_{Y}^{(cr)} = m$, and for $m^2 < 0$,
$\mu_Y$ should be larger
than $\mu_{Y}^{(cr)} = g|m|/2\sqrt{\lambda}$ (the
critical value $g|m|/2\sqrt{\lambda}$ coincides with the mass
of $W$ boson in the vacuum theory with $\mu_Y =0$ and $m^2 <0$).}
   
It is clear that this solution implies that the initial symmetry
$SU(2) \times U(1)_Y \times SO(3)_{\rm rot}$ is spontaneously
broken down to $SO(2)_{\rm rot}$, i.e., the corresponding medium
is anisotropic.
It is noticeable that the
electromagnetic $U(1)_{em}$, with electric charge 
$Q_{em} = I_3 + Y/2$,
is spontaneously broken by the condensate
of $W$ bosons, i.e., electric superconductivity takes place in
this anisotropic medium.

Because the dynamics in this model is under control for small
$g$, $g^{\prime}$ and $\lambda$, the model provides a proof
that the dynamics
with vector condensate is a real thing. Moreover, this dynamics is
very rich. In particular, as was shown in Ref. \citen{Gorbar:2005pi},
there are three types of topologically stable vortices 
in model (\ref{Lagrangian}), which are connected
either with photon field or hypercharge gauge field, or 
with both of them. In Ref. \citen{sigmamodel}, gapless Nambu-Goldstone
excitations and roton-like excitations were revealed in this phase.

It is noticeable that solution (\ref{vacuum}) describes a nonzero
field strength $F_{\mu\nu}^{(a)}$ which corresponds to the presence of
{\it non-abelian} constant ``chromoelectric''-like condensates in the
ground state. Choosing the vacuum with the condensate 
$\langle{W^{(-)}_z}\rangle$ to be real,
we find from Eq. (\ref{vacuum})
\begin{eqnarray}
E_{3}^{(2)} &=& F_{03}^{(2)} = 
g\sqrt{2}\, \langle{W^{(-)}_z}\rangle  \langle{A^{(3)}_0}\rangle =  
gv_0\sqrt{\frac{\mu_Y v_0}{\sqrt{2}g}-\frac{v_0^2}{4}}\,.
\label{chromoel}
\end{eqnarray}
We emphasize that while an abelian constant electric field in
different media always leads to an instability,
\footnote{In metallic and superconducting
media, such an instability is classical in its origin.
In semiconductors and insulators, this instability is
manifested in creation of electron-hole
pairs through a quantum tunneling process.}
non-abelian constant chromoelectric fields do not in many cases. 
For a
discussion of the stability problem for constant non-abelian fields,
see Refs. \citen{Gorbar:2005rx,bw}.  On a
technical side, this difference is connected with that while a vector
potential corresponding to a constant abelian electric field depends
on spatial and/or time coordinates, a constant non-abelian
chromoelectric field is expressed through constant vector potentials,
as takes place in our case, and therefore momentum and energy are good
quantum numbers in the latter.

As we will see below, the dynamics in the gluonic
phase strikingly resembles the dynamics in this renormalizable
model being however much more complicated.

\subsection{Vector condensates, gauge invariance and Gauss law
constraint}
\label{sec4}

Condensates of vector fields are of course not gauge invariant.
What is a gauge invariant description of the physics they
manifest?
In the Higgs phase, as that described in the previous
subsection, the simplest way to achieve this is to use a unitary 
gauge. The important point is
that in the unitary gauge, all
auxiliary (gauge dependent) degrees of freedom are removed.
{\it Therefore in this gauge the vacuum expectations values (VEVs)
$\langle{A^{(a)}_\mu}\rangle$ of vector fields are well-defined
physical quantities.} The unitary gauge in the renormalizable
model in Eq. (\ref{Lagrangian}) is given by the constraint
\begin{equation}
\Phi^T = (0,\phi_0),
\label{unitary}
\end{equation}
where $\phi_0$ is a real field. The unitary gauge in the gluonic
phase in dense QCD, used in Refs. \citen{Gorbar:2005rx,Gorbar:2007vx},
will be described below in Sec. \ref{sec6}.

In the case when a Higgs field is assigned to the
fundamental representation of the gauge group, there exists
a dual, gauge invariant, description in terms of composite
gauge singlet fields. The structure of
these composites for the electroweak theory (and, therefore, for
the present model) is explicitly given in 
Ref. \citen{Dimopoulos:1980hn}. In particular, in this description,
the condensate of a gauge $W^{(\mp)}_z$ field is replaced by
a condensate of a gauge invariant vector composite.  

There is another subtlety in the description of the dynamics with
vector condensates, which is connected with the derivation of
a {\it physical} effective potential, whose minima correspond to
stable or metastable vacua. The point is that
although the gauge symmetry is gone in the unitary gauge,
the theory still has constraints. In fact, it is a system with
second-class constraints, similar to the theory of a free 
massive vector field $A_{\mu}$ described by the Proca Lagrangian
(for a thorough discussion of systems with second-class 
constraints, see Sec. 2.3 in book \citen{GT}).
In such theories, while the Lagrangian formalism can be used
without introducing a gauge, the physical Hamiltonian
is obtained by explicitly resolving the constraints. In 
our case, this implies that 
to obtain the physical effective potential $V_{\rm phys}$
one has to impose the Gauss law constraints on the conventional
effective potential $V$. 

This feature is intimately connected with the presence of
time-like components of vector fields associated with
would be negative norm states.
The Gauss constraints amount to integrating out the 
time-like components. In tree approximation,
this can be done by using their
equations of motion. In particular,
one can show that solution (\ref{vacuum}) is a minimum of the physical 
potential $V_{\rm phys}$.\cite{sigmamodel}
Is it the global minimum and are there
other minima? This question has been recently addressed in
Refs. \citen{Buchel:2006aa,Buchel:2007rx} and we will consider it
in the next subsection.

\subsection{Landscape of vacua}
\label{sec5}

The analysis of minima of the physical potential $V_{\rm phys}$
in this model is quite nontrivial. Indeed, there are 10 physical 
fields in the model,
and the problem is equivalent of studying the geometry of
a ten dimensional hypersurface corresponding to $V_{\rm phys}$. 
Very recently, this analysis has been done for the special case
with the quartic coupling constant $\lambda$
and the mass of the scalar field $\Phi$ chosen to be 
zero.\cite{Buchel:2006aa,Buchel:2007rx} 
This case, retaining richness of
the dynamics, simplifies the analysis of the structure
of the vacuum manifold. This allowed to establish 
that the anisotropic superconducting vacuum
described in Subsec. \ref{sec3} above is the {\it global}
vacuum in the model. Besides that, there are
some metastable vacua. Among them, the
metastable vacua with an abnormal number of Nambu-Goldstone bosons
were identified.  
\footnote{It is the same phenomenon as that found in a non-gauge
relativistic field model at finite density in Ref. \citen{ms}.}   
The $SO(2)$ symmetry of these vacua corresponds to locking gauge,
flavor, and spin degrees of freedom. There are also metastable $SO(3)$
rotationally invariant vacua. Thus, the landscape of vacua in the
model is rich. 
These results encourage studies of the vacuum landscape in
dense QCD (see Ref. \citen{Gorbar:2007vx}).

\section{Gluonic phase in dense two-flavor QCD}
\label{sec6}

Both the chromomagnetic \cite{Huang:2004bg,Huang:2004am} and plasmon 
\cite{Gorbar:2006up} instabilities for the 4-7th gluons
in the 2SC phase at $\bar{\Delta} < \sqrt{2}\delta\mu$
suggest a condensation of these gluon fields. 
Because the chromomagnetic instability develops in the magnetic
channel, it is naturally to expect that 
some spatial components of these fields have
a nonzero VEVs. The latter implies that the rotational symmetry 
should be spontaneously broken. This led to the
suggestion that the underlying dynamics is similar to that in the gauged
$\sigma$ model considered
in Sec. \ref{sec2} above.\cite{hypoth}
Such a solution (the
gluonic phase) was revealed 
in letter \citen{Gorbar:2005rx}. A detailed
description of this phase has been recently given in Ref.
\citen{Gorbar:2007vx}. 

\subsection{Condensates in gluonic phase}
\label{sec7}

At intermediate energy scales of the order of the diquark condensate
$\bar{\Delta} \sim {\cal O}(\mbox{50MeV})$, the analysis of QCD dynamics
is very hard.
Hence the phenomenological Nambu-Jona-Lasinio (NJL) model
plays a prominent role in the analysis in dense quark 
matter.\cite{review,Huang:2003xd,Abuki:2004zk,Ruster:2005jc}
The NJL model is usually regarded as a low-energy
effective theory in which massive gluons are integrated out.
The situation with dense quark matter is however quite different from
that in the vacuum QCD. In our analysis \cite{Gorbar:2005rx,Gorbar:2007vx },
we introduce gluonic degrees of freedom
into the NJL model because in the 2SC/g2SC phase
the gluons of the unbroken $SU(2)_c$ 
subgroup of the color $SU(3)_c$
are left as massless, and, near the critical point 
$\delta\mu=\bar{\Delta}/\sqrt{2}$, plasmons in
the 4-7th gluon channels are also very light \cite{Gorbar:2006up}. 
This yields the gauged NJL model. Our analysis was done in the
framework of the Ginzburg-Landau (GL) approach. Here
I will present a physical picture underlying the
gluonic phase. As we will see, it strikingly resembles that 
corresponding to the gauged $\sigma$ model considered in Sec. \ref{sec2}.

The major ``players'' in the dynamics in the gluonic phase are the
following. The first one is the chemical potential matrix $\hat{\mu}_0$ 
for up and down quarks.
In the $\beta$-equilibrium, it is
\begin{equation}
  \hat{\mu}_0 = \mu  - \mu_e Q_{\rm em} + \mu_8 Q_8\, , 
 \label{mu}
\end{equation}
where 
$Q_{\rm em} \equiv {\rm diag}(2/3,-1/3)_f$ is the electric charge,  
$Q_8 \equiv {\rm diag}(1/3,1/3,-2/3)_c$ is the color charge, and
$\mu$, $\mu_e$ and $\mu_8$
are the quark, electron and color chemical potentials,
respectively
(the baryon chemical potential $\mu_B$ is $\mu_B \equiv 3\mu$
and the mismatch $\delta\mu$ between the Fermi momenta of
the pairing quarks is $\delta\mu = \mu_e/2$).
Here the subscripts $f$ and $c$ mean that the corresponding matrices act in
the flavor and color spaces. The second player is 
the diquark field 
$\rm {\Phi}^\alpha \sim i\bar{\psi}^C\varepsilon \epsilon^\alpha \gamma_5\psi$,
with $(\varepsilon) \equiv \epsilon^{ij}$ and
$(\epsilon^{\alpha}) \equiv \epsilon^{\alpha\beta\gamma}$
being the totally antisymmetric tensors
in the flavor and color spaces, respectively. The last players are
seven gluon fields $A^{(1)}_\mu-A^{(7)}$, which are light near the
critical point $\delta\mu=\bar{\Delta}/\sqrt{2}$.

Let us start from the 2SC/g2SC phase, with subcritical values
of $\delta\mu < \bar{\Delta}/\sqrt{2}$. In this case, the only
condensate is that of the diquark field. 
Without loss of generality, 
the diquark condensate in the 2SC/g2SC phase
can be chosen along the anti-blue direction:
$\langle{\rm{\Phi}^r}\rangle=0, \quad  
\langle{\rm{\Phi}^g}\rangle=0, \quad 
\bar{\Delta}\equiv \langle{\rm{\Phi}^b}\rangle \neq 0$ .

The gap  $\bar{\Delta}$ breaks the color $SU(3)_c$ down to
$SU(2)_c$. The octet of gluons is decomposed with respect to
$SU(2)_c$ as: 
\begin{equation}
  {\bf 8}={\bf 3}\oplus{\bf 2}\oplus\bar{{\bf 2}}\oplus{\bf 1},
  \quad \mbox{i.e.,} \quad
 \{A_\mu^{a}\} = (A_\mu^{(1)},A_\mu^{(2)},A_\mu^{(3)})
 \oplus \phi_\mu \oplus \phi_\mu^* \oplus A_\mu^{(8)} ,
 \quad (a=1,2,\cdots,8).
\end{equation}
Here we defined the complex doublets of the matter (with respect to
the gauge $SU(2)_c$) fields: 
\begin{equation}
  \phi_\mu \equiv
  \left(
  \begin{array}{@{}c@{}} \phi_\mu^r \\[2mm] \phi_\mu^g \end{array}
  \right)
  =
 \frac{1}{\sqrt{2}}
  \left(
  \begin{array}{c}
  A_\mu^{(4)}-iA_\mu^{(5)} \\[2mm] A_\mu^{(6)}-iA_\mu^{(7)}
  \end{array}
  \right) , \qquad
  \phi_\mu^* \equiv
  \left(
  \begin{array}{@{}c@{}} \phi_\mu^{*r} \\[2mm] \phi_\mu^{*g} \end{array}
  \right)
  =
 \frac{1}{\sqrt{2}}
  \left(
  \begin{array}{c}
  A_\mu^{(4)}+iA_\mu^{(5)} \\[2mm] A_\mu^{(6)}+iA_\mu^{(7)}
  \end{array}
  \right) .
\label{phi}
\end{equation}

Because of the chromomagnetic instability, one should expect that
a spatial component of the complex doublet $\phi_\mu$ has
a nonzero VEV for the supercritical values 
of  $\delta\mu > \bar{\Delta}/\sqrt{2}$. 
By using the rotational symmetry $SO(3)_{\rm rot}$, one can take
$\langle{\phi_{z}}\rangle \ne 0$. 
And because of the $SU(2)_c$ symmetry,
without loss of generality, we can choose 
$\langle{A^{(6)}_{z}}\rangle \ne 0$. Acting as a Higgs field,
$\phi_z$ breaks
the $SU(2)_c$ down to nothing. Besides that, 
it also breaks the $SO(3)_{\rm rot}$ down to
$SO(2)_{\rm rot}$. 

The analysis \cite{Gorbar:2005rx,Gorbar:2007vx} 
shows that
there indeed exists a stable
(at least locally) solution in which, besides the 
diquark condensate
$\langle\rm{\Phi}^b\rangle =\bar{\Delta}$, 
there are the following condensates for
gluon fields:
\begin{equation}
\langle\phi_{z}^T\rangle = \frac{1}{\sqrt{2}}
(0, \langle{A^{(6)}_{z}}\rangle \equiv v_{0}^{\prime}), \quad
\langle{A^{(\pm)}_{z}}\rangle \equiv C^{\prime}, \quad
\langle{A^{(3)}_{0}}\rangle \equiv D^{\prime}, 
\label{BCD}
\end{equation}
where $A^{(\pm)}_{\mu} = 1/\sqrt{2}(A^{(1)}_{\mu} \pm iA^{(2)}_{\pm})$. 
These VEVs lead to chromoelectric field strength 
condensates, similar to that in model (\ref{Lagrangian}) (see
Eq. (\ref{chromoel})).

Note that VEVs  
(\ref{BCD}) correspond 
to choosing the following unitary gauge in the gluonic phase:
\begin{equation}
{\rm{\Phi}^T = (0, 0, {\Phi}^b}\equiv\Delta), \quad
\phi_{z}^T = \frac{1}{\sqrt{2}}(0, A^{(6)}_{z})
\label{unitarygl}
\end{equation}
with the fields $\Delta$ and $A^{(6)}_{z}$ being real
(compare with Eq. (\ref{unitary})). 

\subsection{Symmetry breaking structure in the gluonic phase}
\label{sec8}

Comparing Eq. (\ref{BCD})
with Eq. (\ref{vacuum}), one can see that the condensates 
$v_{0}^{\prime}$,
$C^{\prime}$, and $D^{\prime}$ in the gluonic phase play the same
role and the condensates 
$v_0$, $C$, and $D$ in the model (\ref{Lagrangian}).
In particular, as in that case, the $SO(3)_{\rm rot}$ and
the electromagnetic $U(1)_{\rm em}$ 
are spontaneously broken in the gluonic phase, i.e., this
phase describes an anisotropic superconducting medium. Let us 
describe this feature in more detail.

As is well known,\cite{review} in the 2SC/g2SC phase, the generator
of the unbroken $\tilde{U(1)}_{\rm em}$ is 
$\tilde{Q} = Q-\frac{1}{\sqrt{3}}T^8$, where $Q$ is the electric
charge in the vacuum QCD and $T^a$ are the $SU(3)_c$ generators.
The modified baryon charge is $\tilde{\cal B} = 2(\tilde{Q} - I_3)$,
where $I_3$ is the flavor isospin generator.

In the gluonic phase, 
the VEV $v^{\prime}_0 =\langle{A^{(6)}_{z}}\rangle$ 
breaks $SU(2)_c$, but a linear combination
of the generator $T^3$ from the $SU(2)_c$ and $\tilde{Q}$,
$\tilde{\tilde{Q}} = \tilde{Q} - T^3 = Q-\frac{1}{\sqrt{3}}T^8 - T^3$,
determines the unbroken $\tilde{\tilde{U}}(1)_{em}$ (the new baryon
charge is
${\tilde{\tilde{\cal B}}} = 2(\tilde{\tilde{Q}} - I_3)$).
However, because $T^1$ does not commute with $T^3$,
the VEV 
$C^{\prime}=\langle{A^{(1)}_{3}}\rangle$ 
breaks $\tilde{\tilde{U}}_{em}(1)$. The
baryon charge is also broken. The gluonic phase is an anisotropic
superconductor indeed. 

\subsection{Exotic hadrons in gluonic phase}

It is easy to check that the electric charge
$\tilde{\tilde{Q}}_{\rm em}$ and the
baryon number $\tilde{\tilde{\cal B}}$ 
are integer both for all gluons and all quarks
(see Tables I and II in Ref. \citen{Gorbar:2007vx}).
Do they describe hadronic-like excitations? 
We believe that the answer is ``yes". 
\cite{Gorbar:2005rx,Gorbar:2007vx} The point is that because
both the Higgs fields $\rm{\Phi}$ and $\phi_z$ are assigned
to (anti-) fundamental representations of the corresponding
gauge groups ($SU(3)_c$ and $SU(2)_c$, respectively), there 
should exist a dual, gauge invariant (or confinement), description of this 
dynamics.\cite{Dimopoulos:1980hn}. 

Such a description
for the gluonic phase has been recently considered
in Ref. \citen{Gorbar:2007vx}. It was shown there that
all the gluonic and quark fields 
can indeed be replaced by colorless composite ones in the confinement
picture. The flavor quantum numbers of these composite fields
are described by the conventional 
electric and baryon charges $Q_{\rm em}$ and ${\cal B}$. They are 
integer and coincide 
with those the charges $\tilde{\tilde{Q}}_{\rm em}$ and 
${\tilde{\tilde{\cal B}}}$ yield for gluonic and quark fields
(see Tables III and IV in Ref. \citen{Gorbar:2007vx}). 

A very interesting point is that vector hadrons corresponding to
$A^{(\pm)}_\mu$ gluons carry
{\it both} electric and baryon charges: 
$Q_{\rm em} = \pm 1$, ${\cal B} = \pm 2$.
In other words, they are exotic hadrons (in vacuum QCD, bosons carry
of course no baryon charge). The origin
of these exotic quantum numbers is
connected with (anti-) diquarks,
which are constituents of these hadrons (see Table IV in
Ref. \citen{Gorbar:2007vx}).
Indeed, (anti-) diquarks are bosons carrying the baryon charge
$\pm 2/3$ and therefore are exotic themselves. 
This feature has a dramatical consequence for the gluonic 
phase. Since in the Higgs description of 
this phase $A^{(\pm)}$ gluons condense (leading to 
the spontaneous $\tilde{\tilde{U}}(1)_{\rm em}$ breakdown),
we conclude that in the confinement picture
this corresponds to a condensation of {\it exotic} 
charged vector hadrons.
In this regard,
it is appropriate to mention that some authors speculated about
a possibility of a condensation of vector $\rho$ mesons in dense
baryon matter.\cite{rho} 
The dynamics in the gluonic phase 
yields a scenario even with a more unexpected condensation.

\section{Conclusion}

The dynamics in the gluonic phase is extremely rich.
The existence of exotic hadrons there is especially intriguing.
What could be directions for future studies
in gluonic-like phases? It is evident that it 
would be interesting
to consider the spectrum of light collective excitations there.
It is also clear that it would be worth to figure out whether
phases with vector condensates of gluons could exist in dense
matter with three quark flavors. It would be also interesting
to examine a landscape of (stable and metastable) vacua with 
different types of gluon condensates, some of which have
been pointed out in Ref. \citen{Gorbar:2007vx}.
Last but not least, it would
be important to study manifestations of gluonic-like
phases in compact stars.  

\section*{Acknowledgments}

I am grateful to Prof. Taichiro Kugo and Prof. Teiji Kunihiro 
for their warm hospitality during my stay
at Yukawa Institute for Theoretical Physics, Kyoto University. 
Discussions during the YKIS2006 ``New Frontiers in QCD'' were
useful for completing this work. 
The work was supported by the Natural Sciences and Engineering
Research Council of Canada. 


\end{document}